\documentstyle[psfig]{mn}

\newif\ifAMStwofonts

\def\lapp{\ifmmode\stackrel{<}{_{\sim}}\else$\stackrel{<}{_{\sim}}$\fi}
\def\gapp{\ifmmode\stackrel{>}{_{\sim}}\else$\stackrel{>}{_{\sim}}$\fi}
\def\psr{PSR~J2144$-$3933}
\def\ins{RX~J0720.4--3125}
\def\inss{RX~J0806.4--4132}

\title{Radio Observations of Two Isolated Neutron Stars,  \ins{} and \inss{}}
\author[Simon Johnston]
{Simon~Johnston\\
School of Physics, University of Sydney, NSW 2006, Australia.
}
\date{\today}
\pagerange{\pageref{firstpage}--\pageref{lastpage}}
\pubyear{2002}
\begin{document}
\maketitle
\label{firstpage}

\begin{abstract}
Radio observations of two isolated neutron stars, \ins{} and \inss{},
have been made with the Australia Telescope Compact Array at
a frequency of 1.4 GHz.  No continuum emission is detected 
from either object with a 3-$\sigma$ upper limit of 0.2 mJy.
The data were also folded synchronously with the known rotation
periods of 8.4 and 11.4~s respectively. No pulsed emission was
detected. If the pulse duty cycle is small, the upper limit on
pulsed emission can be reduced still further to 0.04 mJy.
The best evidence seems to indicate that the isolated neutron
stars detected by ROSAT all-sky survey are relatively
young objects, born with a very high magnetic field.
\end{abstract}

\begin{keywords}
  neutron stars: individual: \ins{}; \inss{}
\end{keywords}

\section{Introduction}
A new class of neutron stars has emerged with the discovery of the
so-called Isolated Neutron Stars (INS) in the X-ray band with
ROSAT. Their identification as neutron stars is secure as they
have extremely high X-ray to optical flux ratios.
They are isolated in the sense that they have no binary companions
which rules out accretion from the companion
as a mechanism for generating the X-rays.
They are all extremely close, probably only a few hundred parsecs
(at most) from Earth. There is no known radio emission (pulsed or 
unpulsed) associated with these objects although deep
searches have not yet been reported in the literature.
A review of these objects can be found in Treves et al. (2000)\nocite{ttzc00}.

There are 3 possibilities as to the nature of these neutron stars.
First, they might be old stars accreting from the interstellar medium.
These still rotate, albeit slowly, but do not produce `conventional'
radio or X-ray emission. The second possibility is that they are
`standard' (old) neutron stars, with magnetic fields of order 10$^{12}$~G
which pulsate in X-rays and pulsate in the radio but may or may not
be beamed in our direction.
Finally, they may be `magnetars', which are young neutron stars with
extremely high magnetic fields (greater than 10$^{14}$~G).
These have X-ray pulsations
but no radio pulsations. The mechanism for producing X-ray pulsations
is completely different to conventional pulsars; they are powered by
magnetic rather than rotational energy.

When the INS were first discovered a few years ago, it seemed unlikely
that they were standard radio pulsars \cite{hk98}. No radio pulsars at that time
had periods greater than 4 seconds and magnetic fields more
than 10$^{13}$~G. However, in 1999, \psr{} was discovered to have
a period of 8.5 s
(Young, Manchester \& Johnston 1999)\nocite{ymj99}
and Camilo et al. (2000)\nocite{ckl+00} discovered PSR~J1814--1744 
with a long period
and a high magnetic field similar (in parameter space at least)
to the so-called `magnetars'.
\psr{} has a very low dispersion measure, placing
it about 200 pc distant, a similar distance to the known INS.
It therefore now seems more likely that the INS are normal pulsars rather
than some new exotic species.

One way of determining the nature of these objects is detection
of periods and period derivatives. The combination of period and
period derivative naturally yields the magnetic field strength,
which is the key discriminator in the scenarios outlined above.
Four of the 7 known INS have X-ray pulsations with rotation periods
ranging from 5.2 to 22 seconds. Of interest here 
are \ins{} and \inss{}
which have spin periods of 8.4 and 11.4 seconds respectively
\cite{hz02,kkkm02,zhc+02}.
Timing analysis has so far not yielded a period derivative (and hence
magnetic field strength) for these pulsars. Zane et al. (2002) claim
a period derivative for \ins{} but their phase connected
solution has been called into
question by Kaplan et al. (2002). However, the upper limit on
the period derivative seems to rule out the magnetar model.
For \inss{} the upper limit on the
period derivative does not allow discrimination between the models.

The question then is - are there radio pulses and/or
continuum radio emission (e.g. from a wind nebula) from these objects?
If radio pulsations were detected then it becomes relatively easy
to measure a period derivative and hence determine the nature of
these objects.
If detected, they would be among the closest radio pulsars, and likely
the lowest luminosity pulsars known. This would have implications
for the birth rate of pulsars and the low-luminosity cut-off.
It is a non-trivial task, however, to detect these
long period pulsars using conventional search techniques as the
red-noise in the Fourier transform of the time series seriously
reduces the sensitivity. Knowledge of the period {\it a prori}
helps the search process.

\section{Observations and Data Reduction}
Observations of \ins{} and \inss{} were made at the 
Australia Telescope Compact Array (ATCA)
on 2002 December 4 and 5 using the 6A array configuration.
In this configuration the shortest baseline is 627 m and
the longest 5940 m.
In each observation, two continuum frequencies were observed
simultaneously with 32 4-MHz channels across each; all four Stokes
parameters were recorded. The centre frequencies were 1384 and 1704 MHz.
The pointing centres used were RA (J2000) 07$^{h}$ 20$^{m}$ 24.96, Dec (J2000)
--31\degr\ 25\arcmin\ 19.6\arcsec\ and
RA (J2000) 08$^{h}$ 06$^{m}$ 23.47, Dec (J2000) --41\degr\ 22\arcmin\ 02.3\arcsec\
respectively, both $\sim$30\arcsec\ offset from the X-ray location of the INS.
The total integration time was 10~hr per source.
The ATCA is also capable of splitting each correlator cycle into
bins corresponding to different phases of a pulsar's period, and,
in this case, the pulse period divided into 32 time bins.
The flux density scale of the observations was determined by
observations of PKS~B1934--638, while polarization and antenna gains
and phases were calibrated using 3 min observations every 50~min of
PKS~B0614$-$349 for \ins{} and PKS~B0823$-$500 for \inss{}.
As a system test, observations were made of the known 8.5~s
radio pulsar \psr{} using an identical setup to that
described above. The source was observed for 30 min.

Data were edited and calibrated using the MIRIAD package.
At each frequency the field of interest was imaged
using conventional techniques, after first collapsing the phase bins.

The data were also folded to produce a pulse profile. The dispersion
smearing was assumed to be negligable compared to the duration of one
phase bin. The data were folded according to the ephemeris given
in Young et al. (1999) for \psr{}, assuming a barycentric period
of 8.391115 s for \ins{} as given by Kaplan et al. (2002),
and assuming a barycentric period
of 11.3714 s for \inss{} as given in Haberl \& Zavlin (2002).
Note that in the case of \ins{}, if the star has a period derivative
in excess of $3.8\times 10^{-13}$, this would be sufficient to broaden
the pulse profile by more than one bin over the
course of the 10 hr observation.

\section{Results and Discussion}
\subsection{\psr{}}
\begin{figure}
  \centerline{\psfig{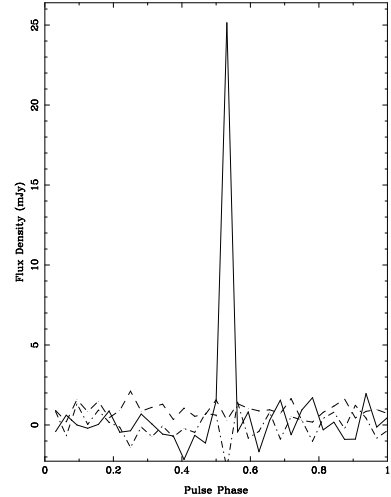}}
  \caption{Pulse profile of \psr{} at 1.4 GHz. Solid line is the total
intensity, dashed line linear polarization and dash-dot circular
polarization.} 
  \label{2144}
\end{figure}
\begin{figure}
  \centerline{\psfig{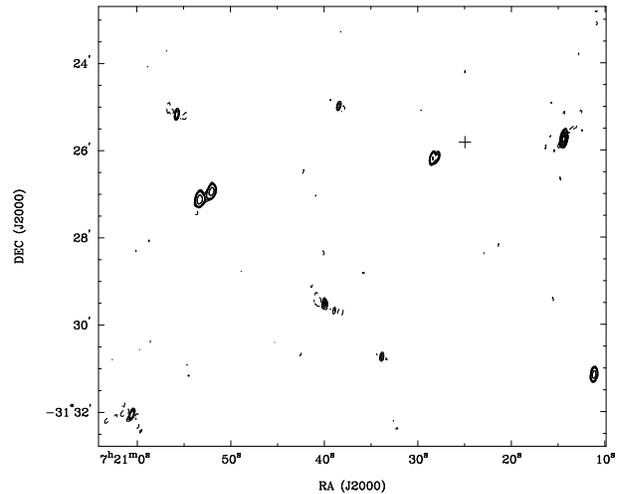}}
  \caption{Radio image of the region around \ins{} at 1.4 GHz.
The 1-$\sigma$ level is 0.06 mJy and the contour
levels are --0.3, 0.4, 0.6, 0.8, 1.0, 2.0 and 3.0 mJy. The position
of \ins{} is marked with a +. The beam is 12\arcsec\ x 4\arcsec\ at
a position angle of --9.5\degr\ measured from north.}
  \label{0720_1}
\end{figure}
\begin{figure}
  \centerline{\psfig{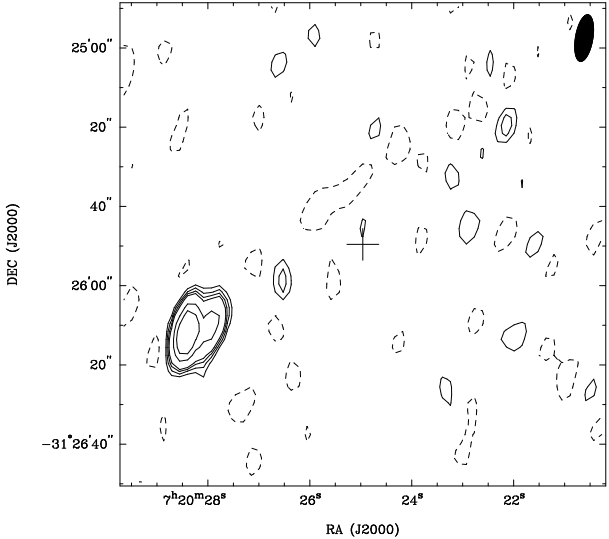}}
  \caption{Radio image of a 2\arcmin\ region centered on \ins{} whose
position is shown by a +. The contour levels are a factor of 2 lower
than shown in Fig~\ref{0720_1}. The beam is shown at the top right.}
  \label{0720_2}
\end{figure}
The pulse profile of \psr{} at 1.4 GHz is shown in Fig.~\ref{2144}. The
pulsar is known to have a short duty cycle and hence its pulse fits
entirely within one phase bin in this instance. The polarization of the
pulsar is low, less than 10\% of the total intensity. This is
consistent with observations at lower frequencies
\cite{mhq98}. The flux density at 1.4 GHz is $\sim$0.8 mJy; the flux density
at 436 MHz is 4 mJy (Young et al. 1999) which yields
a (fairly typical) spectral index of --1.4.
This observation shows the viability
of detecting these long period pulsars using an interferometer.

\subsection{\ins{}}
Continuum images of the region around \ins{} are shown in
Figs.~\ref{0720_1} and \ref{0720_2}. The position of \ins{} as
given by the Chandra observations of Kaplan et al. (2002) is shown
by the + symbol. The extended source seen in Fig.~\ref{0720_2} does
not appear to have an optical counterpart in the Deep Sky Survey
nor in the deeper images of Kulkarni \& van Kerwijk (1998)\nocite{kk98}.
It is likely that it is a background extragalactic source.
No radio emission is detected from \ins{} with a 3-$\sigma$ 
upper limit of 0.2 mJy. A more stringent limit can be placed on
the flux density if we assume
that the pulse is contained within one phase bin (as is the case for
\psr{} above). The upper limit is then reduced by a factor $\sqrt{32}$
to 0.05 mJy.

If \ins{} were a conventional pulsar (i.e. not a magnetar),
limits on the efficiency
of the coupling of the relativistic wind with the interstellar
medium can be determined. Following Gaensler et al. (2000)\nocite{gsf+00},
if a putative wind nebulae was a point source, the efficiency
parameter, $\epsilon$ is given by
\begin{equation}
\epsilon = \frac{d^2\,\, S}{2.1\times 10^2\,\,\dot{E}_{31}}
\end{equation}
where $d$ is the distance to the source in kpc, $S$ is the flux
density upper limit in mJy and $\dot{E}$ is the spin-down energy in
units of $10^{31}$ erg~s$^{-1}$.
For \ins{}, therefore, with $d=0.2$~kpc and $\dot{E} < 2\times 10^{31}$
erg~s$^{-1}$, $\epsilon$
would be $<5\times 10^{-5}$, a value not atypical
of other middle-aged pulsars. Therefore we cannot conclude from the lack
of a radio nebula that this is not a neutron star powered by spin-down.

Unfortunately, the lack of radio emission from \ins{} does little
to help our understanding of the nature of this source.
The beaming fraction of these long period pulsars is expected to be
less than 10\% on theoretical grounds \cite{tm98} and the
radio luminosity is low, as seen in \psr{}. The upper limit to
pulsed emission from this source is a factor 10 better than for
\psr{} and, assuming a similar distance of 180 pc makes it unlikely
that it is beaming in our direction but is below the sensitivity limit.
It therefore remains likely that this is a conventional pulsar in 
which the radio beams do not intersect Earth \cite{bj98}.
The question then remains
as to whether it is an old pulsar, with an age similar to that
of \psr{} or a much younger pulsar. The former implies that the age derived 
from the blackbody fits to the X-ray spectrum was incorrect whereas
the latter implies that it must have been born with a very long period,
unlike the bulk of the known pulsars.

\subsection{\inss{}}
\begin{figure}
  \centerline{\psfig{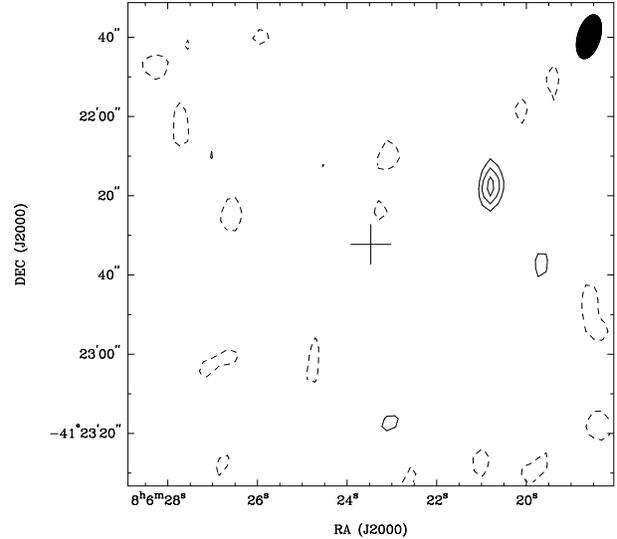}}
  \caption{Radio image of a 2\arcmin\ region centered on \inss{} whose
position is shown by a +. The 1-$\sigma$ level is 0.05 mJy and the contour
levels are --0.3, 0.4, 0.6, 0.8, 1.0, 2.0 and 3.0 mJy.
The beam, shown on the top right, has dimensions of 11\arcsec\ x 6\arcsec\ at
a position angle of --16.2\degr\ measured from north.}
  \label{0806_1}
\end{figure}
A continuum image of a 2\arcmin\ region around \inss{} is shown in
Fig.~\ref{0806_1}. The position of \inss{} as
given by the XMM-Newton observations of Haberl \& Zavlin (2002)\nocite{hz02}
is shown. The area covered by the + symbol is 3 times larger than the
error box given by Haberl \& Zavlin (2002)\nocite{hz02}.
Only 1 source is seen in the field. It has a flux density of 0.4 mJy and is
offset from the position of \inss{} by 34\arcsec.
A search for pulsations at the known period failed to reveal any
significant detection. Assuming the pulsed flux is contained within
one phase bin implies a lower limit on the pulsed flux
of 0.05 mJy.

The pulsed fraction in X-rays is very low for \inss{}, only 6\%.
Haberl \& Zavlin (2002)\nocite{hz02} speculate that perhaps the neutron
star has its magnetic axis and rotational axis nearly aligned.
If this is the case, then radio emission might also be expected.
The lack of a detection in the radio either implies that this proposition
is not correct, that the radio pulsar is an order of magnitude less
luminous than \psr{}, already one of the lowest luminosity pulsars known
or perhaps that it has crossed the death-line and is incapable of
producing radio emission.

\subsection{Speculation on the nature of the INS}
\begin{table*}
\begin{tabular}{lccccc}
Pulsar & \multicolumn{1}{c}{Period} &
\multicolumn{1}{c}{Distance} & \multicolumn{1}{c}{$\dot{E}$} &
\multicolumn{1}{c}{X-ray} & \multicolumn{1}{c}{Radio} \\
& \multicolumn{1}{c}{(s)} &
\multicolumn{1}{c}{(pc)} & \multicolumn{1}{c}{($\times 10^{31}$erg/s)} \\
\hline
Geminga      & 0.237  & 157$^a$ & 3000   & $\surd$  & $\times$ \\
B1929+10     & 0.226  & 331$^a$ & 390    & $\surd$  & $\surd$ \\
B0950+08     & 0.253  & 262$^a$ & 56     & $\surd$  & $\surd$ \\
B0823+26     &  0.531 & 375$^b$ & 45     & $\surd$  & $\surd$ \\
B1133+16     &  1.19  & 350$^a$ & 8.5    & $\times$ & $\surd$ \\
J0108--1431  & 0.808  & 128$^b$ &  0.58  & $\times$ & $\surd$ \\
J2307+2225   &  0.536 & 385$^b$ & 0.22   & $\times$ & $\surd$ \\
J2144--3933  &  8.51  & 179$^b$ & 0.0032 & $\times$ & $\surd$ \\
\ins{}       &  8.39  &     & $<$2.4 & $\surd$  & $\times$ \\
\inss{}      & 11.37  &     &        & $\surd$  & $\times$ \\
\hline
\multicolumn{6}{c}{($a$) parallax measurement,
($b$) estimated from the dispersion measure}
\end{tabular}
\end{table*}
Table~1 lists the 
7 known radio pulsars and the X-ray pulsar Geminga
within 400 pc of Earth (excluding the millisecond pulsars) ranked
in descending order of $\dot{E}$ along
with the two INS observed here.
Of these, the four with the highest $\dot{E}$ have been detected
in the X-ray band, essentially due to the sensitivity of the ROSAT all-sky
survey.  Nearby, spin-powered pulsars, must have
$\dot{E}$ greater than $\sim$5$\times 10^{31}$ erg~s$^{-1}$ in
order to be detected.
Only PSR B1133+16 of the non-detected pulsars has an $\dot{E}$
slightly in excess of this value.
\psr{} has an $\dot{E}$ which is three orders of magnitude less than
this and therefore not only is the lack of X-ray emission from this
pulsar not surprising, but it also seems unlikely that the INS
share the same characteristics.

It therefore seems most likely that the
INS will turn out to have $\dot{E}$ in excess of
5$\times 10^{31}$ erg~s$^{-1}$, consistent with the inferred $\dot{E}$
of RX~J1856.5--3754 \cite{kk01,kka02} and marginally consistent with the
upper limit derived by Kaplan et al. (2002) for \ins{}.
For \inss{}, the combination of an $\dot{E}$ of 5$\times 10^{31}$ erg~s$^{-1}$
and the spin period of 11.4~s would imply a high value of the period derivative
($\sim 2\times 10^{-12}$) and a magnetic field in the magnetar range
($\sim 2\times 10^{14}$ G) and even more extreme parameters for the 22.7 s
INS RX~J0420.0--5022. These could then be magnetars where the energy
source derives from the magnetic field rather than the 
spin-down. In summary, the INS appear to be one of 
the manifestations of neutron stars born with high magnetic fields.
Their number suggests they must form a substantial fraction of the 
population of neutron stars in the Galaxy but that their radio 
emission may be inhibited.

\section*{Acknowledgements}
The Australia Telescope is funded by the Commonwealth of 
Australia for operation as a National Facility managed by the CSIRO.
I thank Warwick Wilson for his efforts beyond the call of duty
in programming the correlator to ensure
maximum efficiency for these long period pulsars.

\bibliography{modrefs,psrrefs,crossrefs}

\begin{thebibliography}{{{Young}, {Manchester} \& {Johnston} }{1999}}

\bibitem[\protect\citename{Brazier \& Johnston }{1999}]{bj98}
Brazier~K. T.~S., Johnston~S., 1999, MNRAS, 305, 671

\bibitem[\protect\citename{Camilo {\rm et~al. }}{2000}]{ckl+00}
Camilo~F., Kaspi~V.~M., Lyne~A.~G., Manchester~R.~N., Bell~J.~F., D'Amico~N.,
  McKay~N.~P.~F., Crawford~F., 2000, ApJ, 541, 367

\bibitem[\protect\citename{Gaensler {\rm et~al. }}{2000}]{gsf+00}
Gaensler~B., Stappers~B., Frail~D., Moffett~D., Johnston~S., Chatterjee~S.,
  2000, MNRAS, 318, 58

\bibitem[\protect\citename{Haberl \& Zavlin }{2002}]{hz02}
Haberl~F., Zavlin~V.~E., 2002, A\&A, 391, 571

\bibitem[\protect\citename{Heyl \& Kulkarni }{1998}]{hk98}
Heyl~J.~S., Kulkarni~S.~R., 1998, ApJ, 506, L61

\bibitem[\protect\citename{Kaplan {\rm et~al. }}{2002}]{kkkm02}
Kaplan~D.~L., Kulkarni~S.~R., van Kerkwijk~M.~H., Marshall~H.~L., 2002, ApJ,
  570, L79

\bibitem[\protect\citename{Kaplan, van Kerkwijk \& Anderson }{2002}]{kka02}
Kaplan~D.~L., van Kerkwijk~M.~H., Anderson~J., 2002, ApJ, 571, 447

\bibitem[\protect\citename{Kulkarni \& van Kerkwijk }{1998}]{kk98}
Kulkarni~S.~R., van Kerkwijk~M.~H., 1998, ApJ, 507, L49

\bibitem[\protect\citename{Manchester, Han \& Qiao }{1998}]{mhq98}
Manchester~R.~N., Han~J.~L., Qiao~G.~J., 1998, MNRAS, 295, 280

\bibitem[\protect\citename{Tauris \& Manchester }{1998}]{tm98}
Tauris~T.~M., Manchester~R.~N., 1998, MNRAS, 298, 625

\bibitem[\protect\citename{Treves {\rm et~al. }}{2000}]{ttzc00}
Treves~A., Turolla~R., Zane~S., Colpi~M., 2000, PASP, 112, 297

\bibitem[\protect\citename{van Kerkwijk \& Kulkarni }{2001}]{kk01}
van Kerkwijk~M.~H., Kulkarni~S.~R., 2001, A\&A, 380, 221

\bibitem[\protect\citename{{Young}, {Manchester} \& {Johnston} }{1999}]{ymj99}
{Young}~M.~D., {Manchester}~R.~N., {Johnston}~S., 1999, Nat, 400, 848

\bibitem[\protect\citename{Zane {\rm et~al. }}{2002}]{zhc+02}
Zane~S., Haberl~F., Cropper~M., Zavlin~V.~E., Lumb~D., Sembay~S., Motch~C.,
  2002, MNRAS, 334, 345

\end{thebibliography}
\bibliographystyle{mn}
\label{lastpage}
\end{document}